\documentclass[reprint,amssymb, amsmath, aip]{revtex4-1}
\usepackage{graphicx}
\usepackage{color}
\usepackage{epsfig}
\usepackage{bm}%
\usepackage[colorlinks=true,linkcolor=blue]{hyperref}%
\expandafter\ifx\csname package@font\endcsname\relax\else
 \expandafter\expandafter
 \expandafter\usepackage
 \expandafter\expandafter
 \expandafter{\csname package@font\endcsname}%
\fi
\begin{document}

\title{Dusty Plasma Experimental (DPEx) device for complex plasma experiments with flow}%
\author{S.Jaiswal}%
\email{surabhijaiswal73@gmail.com}
\author{P.Bandyopadhyay}
\author{A.Sen}
\affiliation{Institute For Plasma Research, Bhat, Gandhinagar,Gujarat, India, 382428}%
\date{\today}
%**************************************************************
%#####################################################################################                  ABSTRACT
%************************************************************************************************
\begin{abstract}
A versatile table-top dusty plasma experimental (DPEx) device to study flow induced excitations of linear and nonlinear waves/structures in a complex plasma is presented. In this {$\Pi$}-shaped apparatus a DC glow discharge plasma is produced between a disc shaped anode and a grounded long cathode {tray} by applying a high voltage DC in the background of a neutral gas {(Argon)} and subsequently a dusty plasma is created by introducing micron sized dust particles that get charged and levitated in the sheath region. A flow of the dust particles is induced in a controlled manner by adjusting the pumping speed and the gas flow rate into the device. A full characterisation of the plasma, using  Langmuir and emissive probe data, and that of the dusty plasma using particle tracking data with the help of an idl based (super) Particle Identification and Tracking (sPIT) code is reported. Experimental results on the variation of the dust flow velocity as a function of the neutral pressure and the gas flow rate are given. {The neutral drag force acting on the particles and the Epstein coefficient are estimated from the initial acceleration of the particles}. The potential experimental capabilities of the device for conducting fundamental studies of flow induced instabilities are discussed. 
\end{abstract}
%%%%%%%%%%%%%%%%%%%%%%%%%%%%%%%%
\maketitle
%%%%%%%%%%%%%%%%%%%%%%%%%%%%%%%%%%%%%%%%%%%%%%%%%%%%%%%%%%%%%%%%%%%%%%%%%%%%%%%%%%%%%       INTRODUCTION
%%%%%%%%%%%%%%%%%%%%%%%%%%%%%%%%%%%%%%%%%%%%%%%%%%%%%%%%%%%%%%%%%%%
\section{Introduction} 
A dusty plasma is characterised by the presence of micron or sub-micron sized macro particles in a normal plasma consisting of electrons, positive and/or negative ions as well as neutrals particles. The presence of the macro particles which usually get negatively charged\cite{barken} by absorbing more electrons than ions from the plasma adds much richness to the collective dynamics of the system. Such plasmas are therefore often called {\it complex} plasmas and the dust component, in particular, displays a variety of equilibrium states. Dusty plasmas occur in a variety of natural situations such as planetary rings, comet tails, interplanetary media and interstellar clouds\cite{goertz,nakamura1}.  The presence of the dust component provides an additional degree of freedom leading to new linear collective modes and their associated nonlinear structures. A large number of studies in the past have been devoted to the investigation of such excitations, e.g. Dust Acoustic Waves (DAW)\cite{rao,barken2}, Dust Lattice Waves (DLW)\cite{melands}, Transverse Shear Waves (TSWs)\cite{kaw,pramanik} and Dust Ion Acoustic Waves (DIAW)\cite{shukla1}, as well as nonlinear waves and structures like Dust Acoustic Solitary Wave (DASW)\cite{shukla2,pintu1}, Dust Ion Acoustic Solitary Waves (DIASW)\cite{washimi,nakamura3}, Dust Acoustic Shocks (DAS)\cite{hamid} and  Ion Acoustic Shocks (IAS)\cite{nakamura4} etc.  Due to the heavy mass of the dust particles the waves associated with their dynamics are typically in the low frequency range and propagate slowly ($\sim$ few $cms/sec$). \par

The slow speed of wave propagation and the large size of the particles allows for a visual detection of the particles as well as the wave motion and is a major diagnostic advantage in the experimental study of dusty plasmas. Typically a combination of lasers (to illuminate the dust particles) and cameras (to record the dynamics of the dust) is used to study wave excitations of dusty plasmas created in glass tube devices.  Examples of some present devices devoted to wave experiments in dusty plasmas include the PK-4\cite{pk4} device - a {$\Pi$}-shaped glass tube that has been used both in the laboratory and under micro-gravity conditions. The device can sustain a combination of RF and DC induced plasma discharges and has been used to study  a rich variety of experiments on dust crystallisation as well as on wave propagation in a dusty plasma. A limited number of experiments have also been done on flow induced instabilities. Nakamura \textit{et al.}\cite{nakamura2}  and Sumita \textit{et al.}\cite{sumita} have carried out wave propagation experiments in a linear glass device and have reported observations on linear dust acoustic waves as well as nonlinear shocks. There is a provision in one of their set-ups (YCOPEX)\cite{nakamura2} to  tilt the experimental device by adjusting a flexible jack placed under one of the edges of the tube. This can result in a flow of the dust particles from a region of higher potential energy to a region of lower potential energy. This flow has been used to create bow shock formations\cite{saitu} in a dusty plasma. The mechanically generated gravity induced flow is somewhat limited in its range and also not easily amenable to control in a precise manner. 
One of the disadvantages of a linear device is that the axial ports at the two ends can not be used for optical observations due to obstructions from the electrodes and therefore there is a restriction on the overall view of the complete dust cloud. To overcome this difficulty we have built a new {$\Pi$}-shaped Dusty Plasma Experimental (DPEx) device at the Institute for Plasma Research which has several in-built facilities for studying flow induced instabilities in a dusty plasma. Apart from permitting easy viewing along the axis, the device generates dust flow through a controlled flow of the neutral gas and the vacuum pumping speed.  
The particle flow can be changed very accurately from a few $mm/sec$ to several $cm/sec$. \par
In this paper we provide a detailed technical and physical description of this device and also present experimental results that provide a complete characterisation of the plasma, the dusty plasma and the dust flow. 
\section{Description of experimental device}\label{sec:theory}
%%%%%%%%%%%%%%%%%%%%%%%%%%
\begin{figure}[ht]
\includegraphics[width=0.5\textwidth]{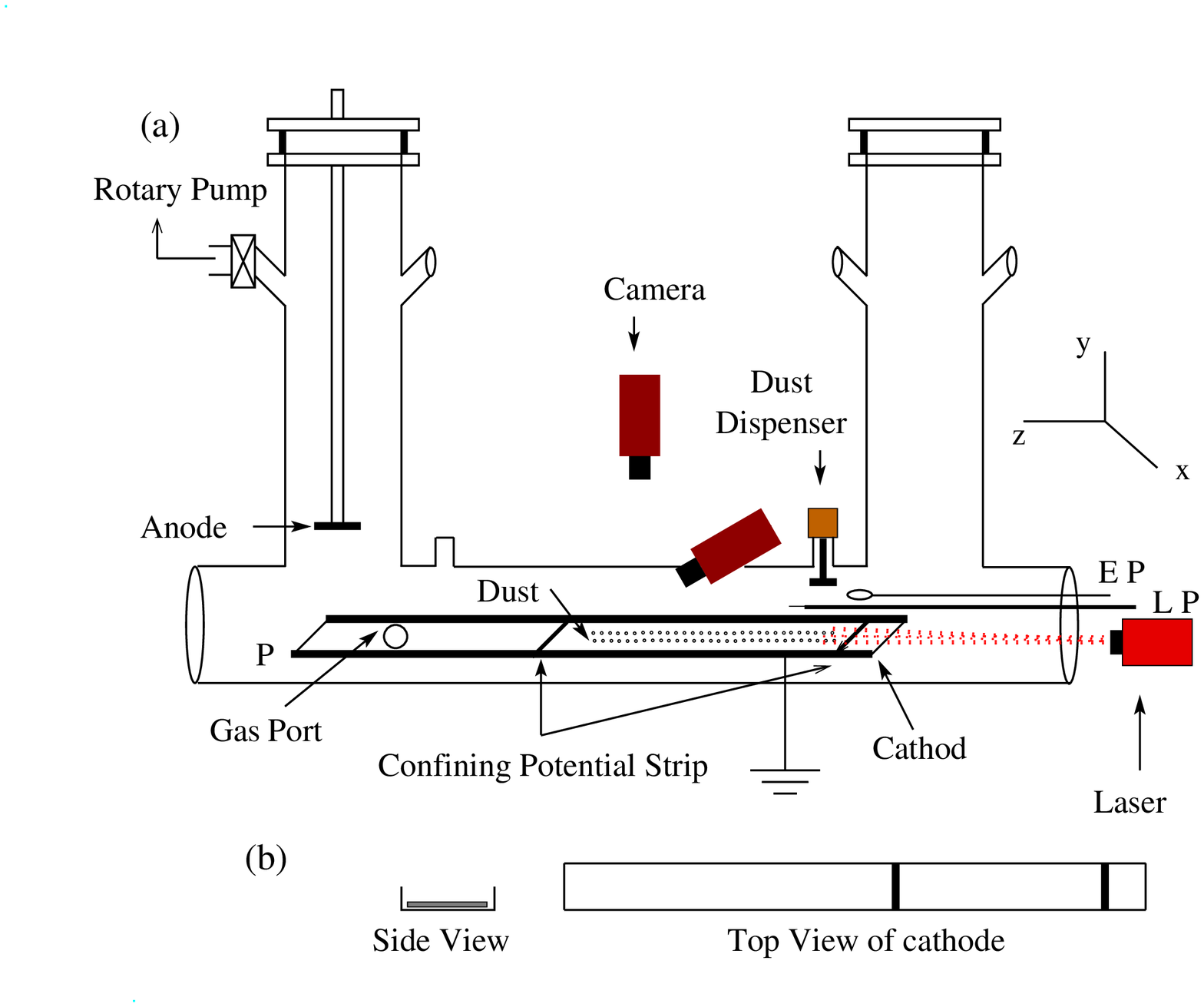}
\caption{(a) Schematic diagram of Dusty Plasma Experimental (DPEx) setup. LP: Langmuir Probe, EP: Emissive Probe, P: grounded cathode {tray}. {(b) Cross sectional  views of the grounded cathode tray.} }
\label{fig:fig1} 
\end{figure}
%%%%%%%%%%%%%%%%%%%%%%%%%   
  A schematic diagram of the Dusty Plasma Experimental (DPEx) device and associated diagnostics is shown in Fig.~\ref{fig:fig1}(a). The vacuum vessel consists of a {$\Pi$}-shaped pyrex glass tube whose arms are each of 30 cms length and have an inner diameter of 8 cms. The connecting tube is of the same diameter and is 65 cms long.  The $\Pi$-tube has several axial and radial ports for functional access. {It may be mentioned here that the geometry of our device and its design philosophy have been inspired a great deal by those of the PK-4 \cite{pk4} device. However there are essential differences between the two devices. The principal differences are in the volume of the device (ours is larger), the electrode geometries and the plasma production mechanisms. In addition, the pumping and gas feeding ports are connected in a different manner in DPEx so as to facilitate precise control of particle flows for experimental investigations of flow induced instabilities.} \par 
The chamber is evacuated by a rotary pump connected to one of the radial ports of the left arm of the tube. Controlled amounts of neutral gas (e.g. Argon) can be fed into the glass tube through a mass flow controller which is interfaced with a computer through a RS-232 cable. A stainless steel circular disc of $4$ cms diameter and $5$ mms thickness, hung from the axial port of the left arm of the vacuum vessel serves as an anode while a stainless steel {tray} of 2 mms thickness, 6.1 cms width and 40 cms length placed inside the connecting tube along the axis serves as the grounded cathode. {The two sides of the cathode plate are bent 1.5 cms to give it the form of a tray that provides radial confinement of the dust particles. A couple of stainless steel strips of dimension 6 cms $\times$ 1.2 cms $\times$ 0.5 cms are placed 12 cms apart on the cathode tray  to provide axial confinement to the particles as shown in Fig \ref{fig:fig1}(b)}. To introduce dust particles into the device a dust dispenser is mounted on a radial port of the connecting tube near the right arm of the {$\Pi$}-tube. The dispenser utilizes an electro-magnetic induction device that is wound using an enamelled copper wire. Mono-dispersive, spherical Melamine-formaldehyde (MF) particles of $9.19\pm0.1 \mu$m diameter and mass density of 1.51 g/cm$^3$ are loaded in a cap of the dust dispenser. The cap is covered with a stainless steel mesh grid having an inter-grid spacing of 11 $\mu$m so that the particles can come out easily. Several spherical steel balls of $1$ mm of diameter are also loaded in the cap along with the particles to shatter the powder into individual particles if they coagulate together. The dust dispenser is connected to a trigger circuit which is interfaced with the same computer and operated using a lab view based software.\par 
To operate the device for experiments on dusty plasmas, the chamber is initially pumped down to its base pressure of $\sim10^{-3}$ mbar. The working gas (Argon) is flushed several times and pumped down to its base pressure. Finally, the working pressure is set to 0.1-0.2 mbar by adjusting the conductance of the pump and the gas flow rate. A high voltage (300 volt--500 volt) is applied between the anode and the grounded cathode at the working pressure to produce a plasma with the help of a Direct Current power supply (2KV, 500mA). Soon thereafter, dust particles are introduced into the plasma by shaking the dust dispenser. Depending on the experimental requirements, one can control the number of particles by tuning the trigger signal. The particles get negatively charged  when they enter into the plasma by collecting more electrons than ions and are trapped in the plasma sheath boundary region above the grounded cathode. The vertical component of the cathode sheath electric field provides the necessary electrostatic force to the particles to levitate them against the gravitational force. The radial and the axial sheath electric fields are responsible for the radial and axial confinement of the dust particles against their mutual coulomb repulsive forces. By adjusting the pumping speed and the gas flow rate, a steady-state equilibrium dust cloud can be formed over the cathode in between the stainless steel strips. {When the gas feed rate is decreased the dust particles flow towards the pump and are constrained by the sheath potential of the strip which is placed approximately in the middle of the cathode. Almost all the particles return to their original position (from where they have started their journey) when the gas flow rate is set to its original value. If the change in the flow rate is high enough then some of these flowing particles overcome the potential barrier and flow over the strip and fall down on the left edge of the glass tube where the cathode ends. After a few experimental shots if the number of particles has substantially reduced then a fresh batch of dust particles 
is introduced.}\par
The particle-cloud is illuminated by a combination of red laser light (650 nm, 50 mW, $\sim$ 2mm beam diameter) and a line generator. The laser illuminates the axial length of the connecting tube so that we can see the dust cloud over its entire length. The Mie-scattered light from the dust particles is captured by a couple of CCD cameras and the images recording the dynamics of the particles are stored into a high speed computer. The high speed camera (60fps, 1MP) is placed at an angle ${15^\circ}$ with the y-axis and the high resolution camera (15fps, 4MP) is placed exactly perpendicular to the dust cloud. The speed of the cameras can be changed by sacrificing the resolution.
\section{Characterisation of a Plasma}
We now describe some initial experimental runs carried out on the device for basic measurements to characterize the  plasma. After striking the discharge, we have made measurements of the plasma properties in the absence of dust particles by using a couple of electrostatic probes, namely a Langmuir probe and an emissive probe. {In principle, the plasma characteristics can change in the presence of dust particles. However when the number density of dust particles is quite low compared to the electron density  the floating and plasma potentials remain relatively unaffected. For most of our experimental conditions the typical dust particle density is of the order of $\sim 10^9$/m$^{3}$ which is much smaller than the typical electron density values of $\sim 10^{15}$/m$^{3}$. Hence the measured plasma characteristics in the absence of dust can be reliably used while studying the dynamics of the dusty plasma.}
Both the probes are introduced into the plasma from the right sided axial port of the connecting tube shown in Fig.~\ref{fig:fig1}. The Langmuir probe is used to estimate the plasma density, the electron temperature, the floating and the plasma potentials by using its $V-I$ characteristics. The emissive probe is used to measure the floating and plasma potentials and hence the electron temperature. The details are given in the following subsections.
%################# ################################################# 
\subsection{Langmuir probe measurements:}
\begin{figure}[ht]
\includegraphics[width=0.45\textwidth]{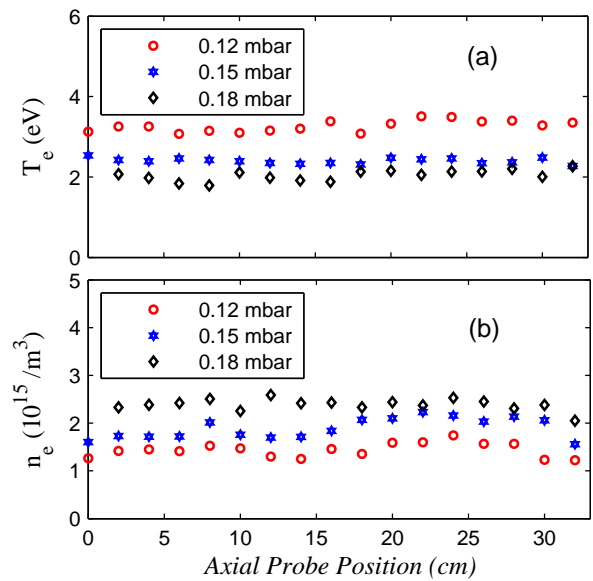}
\caption{Axial profile of (a) electron temperature and (b) plasma density for different values of working pressure. The measurement
errors are within $\pm$ 5$\%$.}
\label{fig:fig8} 
\end{figure}

A cylindrical Langmuir probe is used to measure the plasma parameters, \textit{e.g.} plasma density, electron temperature, floating and plasma potentials over a wide range of discharge parameters. Initially, the floating potential ($V_f$) is measured directly for a given set of discharge parameters and then a ramp voltage of amplitude $V_f-100$ volt to $V_f+100$ volt  is applied to the probe to get a complete information of the ion saturation region to electron saturation region in the probe current. The electron temperature is estimated from the inverse of the slope of the transition region of V-I characteristics curve and the plasma density is calculated from the ion saturation region with the help of the estimated electron temperature. It is observed that the electron temperature changes from 2eV to 4eV whereas the plasma density is of the order of  {$10^{15}$/m$^3$} for a range of discharge voltages between 250--350 volt and working pressures between 0.1--0.2 mbar. The Langmuir probe is then scanned axially to get the temperature and density profiles along the axis of the tube. The variation of the density and the temperature along the axis of the tube are shown in Fig.~\ref{fig:fig8} for three different values of working pressures, namely, P=0.12, 0.15 and 0.18 mbar for a given value of discharge voltage ($V_d=310$ volt).   
Fig.~\ref{fig:fig8} also shows that the density increases and the temperature decreases with the increase of neutral pressure.
%%%%%%%%%%%%%%%%%%%%%%%%%%%
%%%%%%%%%%%%%%%%%%%%%%%%%%%
\subsection{Emissive probe measurements:}
An emissive probe of diameter 0.125 mms and length 4 mms is used to measure the plasma and floating potentials. For a particular position and discharge condition ($V_d=290$ volt and $P=0.12$ mbar), the floating potential and the plasma potential are measured by following three different techniques, i.e., a) Floating method, b) Separation point method and c) Inflection point method (for details see ref \cite{sheenan}). It is found that the measured plasma potential, using the floating method is 269 volt at a distance of 15 cm axially away from the anode. The separation point and the inflection point methods also give the plasma potential to be $269\pm5$ volt for the same position and discharge condition, which is very close to that of the floating method. Hence, the easiest technique, namely the floating method is chosen to measure the floating potential and plasma potential profiles.
%%%%%%%%%%%%%%%%%%%%%%%%%%%%%%%%
\begin{figure}[ht]
\includegraphics[width=0.50\textwidth]{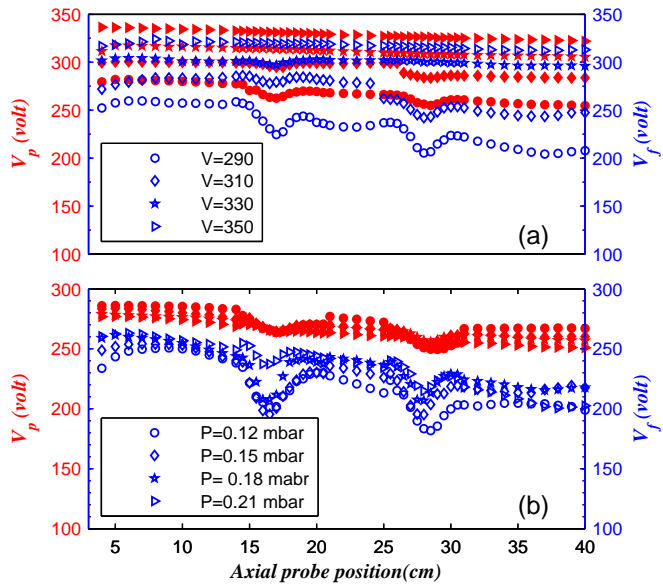}
\caption{Axial profile of plasma,$V_p$ (closed symbols) and floating $V_f$ (open symbol) potential for different (a) discharge voltages at $P=0.120$ mbar and (b) working pressures at $V_d=310$ volt. The measurement
errors are within $\pm$ 5$\%$.}
\label{fig:fig4} 
\end{figure}
%%%%%%%%%%%%%%%%%%%%%%%%%%%%%%
Fig.~\ref{fig:fig4}(a) shows the variation of floating (open symbols) and plasma (solid symbols) potentials along the axial direction at a particular value of the neutral gas pressure (P=0.12 mbar) in the presence of the confining strips.  Four different symbols represent different values of discharge voltages ($V_d=290, 310, 330$ and $350$ volt).  The first point represents the floating/plasma potential when the probe is kept just under the anode whereas the last point represents the same at the position near the right edge of the cathode. It is seen that both the potentials have a small negative gradient when the probe is scanned away from the anode. The difference of plasma potential ($V_p$) and the floating potential ($V_f$) gives the electron temperature ($V_p-V_f=5.2\times kT_e/e$, for argon\cite{merlino}). It is to be noted that the plasma/floating potential decreases with the decrease of the discharge voltage at a given value of pressure. Furthermore, the difference between the plasma potential and the floating potential increases when the discharge voltage is decreased, which indicates that the electron temperature increases with the decrease of discharge voltage. This trend is also seen in the Langmuir probe data (not shown here). Additionally, a couple of wells are observed in the potential profiles just above the confining potential strips. It is also clear from Fig.~\ref{fig:fig4}(a) that the depth of the potential wells (plasma as well as floating potential) decreases with the increase of discharge voltage. Beyond $V_d=350$ volt the effect of the confining potential strips becomes insignificant. {In all the cases, the emissive probe is scanned axially 2.5 cm above the cathode (or 2 cm above the strip, as the strip thickness is 0.5cm). The Debye length (estimated from the electron density and temperature) for the discharge condition, $V_d$=350 volt and 0.120 mbar pressure, comes out to be 0.21 mm whereas it is 0.48 mm for $V_d=290$ volt and $P=0.120 $ mbar. The sheath thickness for these two specific discharge conditions are measured experimentally by examining the snap shots of the motion of the poly-dispersive particles. Inside the plasma, most of the particles are found to be confined by the same confining potential strip which was used in emissive probe measurements. Some of the lighter particles which are levitated in the top most layer are found to flow over the potential hill. By capturing this image, we are then able to measure the sheath thickness very precisely knowing the height of the potential hill from the cathode. In the above two extreme cases the sheath thickness comes out to be 2.21 cms for case-I ($V_d$=350 volt and $P=0.120$ mbar) whereas 2.75 cm for case-II ($V_d=290$ volt and $P=0.120$ mbar). This estimation of the sheath thickness agrees well with the emissive probe measurements (shown in Fig.~\ref{fig:fig4}(a)). It is clear from these measurements that the probe scans over the sheath in case-I whereas it scans through the sheath in case-II.  As a result we find  couple of dips in plasma/floating potential profiles shown  in Fig.~\ref{fig:fig4}(a) for the 2nd case (open and closed circles) whereas no dips are found in the first case (open and closed triangles).} \par
The axial profiles of the floating potential (indicated by open symbols) and the plasma potential (indicated by closed symbols) at a constant discharge voltage, $V_d=310$ volt, are shown in Fig.~\ref{fig:fig4}(b). The different symbols represent different values of background pressures (P=0.12, 0.15, 0.18 and 0.21 mbar). The profiles plotted in this figure show that there is a small gradient in plasma/floating potential while going away from the anode similar to Fig.~\ref{fig:fig4}(a). {We believe this is caused by a small negative gradient in the plasma density along the axis. This is not very clear from Fig.~\ref{fig:fig8} since the axial profiles of density and temperature shown in it are plotted only upto $32$ cm, whereas the potential profiles in Fig.~\ref{fig:fig4} are plotted (from one end to the other end of the cathode) upto 40 cm. We have
independently measured the density over this entire length, using a Langmuir probe and found the value to vary by
$~5\times 10^{14}$/m$^3$ over a distance of 40 cms at a discharge voltage of 310 {volt} and {P=0.12} mbar and thereby ascertained the existence of a density gradient. Our conjecture for the cause of the gradient in the plasma/floating potential is also consistent with the trend seen in the ion saturation current which decreases gradually when the probe is moved away from the anode}. It is also to be noted that both the plasma and floating potentials decrease with the increase of neutral gas pressure although the change is not significant. Similar to Fig.~\ref{fig:fig4}(a), a couple of potential wells are also found in the potential profiles. These wells are more prominent in case of floating potential profiles compared to the plasma potential profiles.  The depth of the wells decreases when the working pressure is increased which implies that the strength of the confining potential decreases with the increase of pressure.\par
Hence, it can be deduced from Fig.~\ref{fig:fig4}(a) and Fig.~\ref{fig:fig4}(b), that lower pressure and/or lower discharge voltages provide a favourable condition for confinement of dust particles in the axial direction at a particular height from the cathode.
\section{Characterisation of a dusty plasma}
\begin{figure}[ht]
\includegraphics[width=0.45\textwidth]{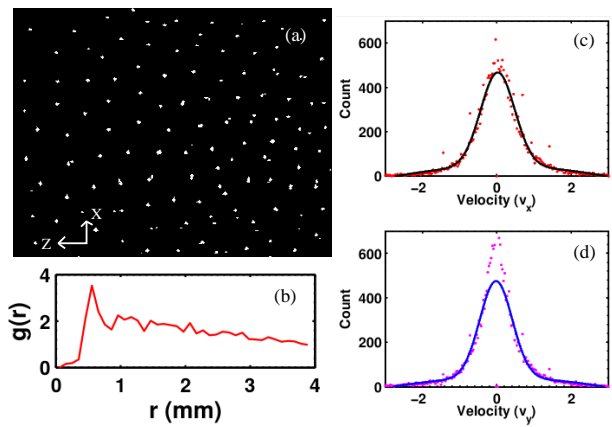}
\caption{(a) Top view of micro-particles forming a 2D crystal lattice, (b) pair correlation function calculated from Fig~\ref{fig:fig5}(a). (c) velocity distribution function along x and (d) velocity distribution function along y.} 
\label{fig:fig5} 
\end{figure}

To characterise the dusty plasma, we have carried out some initial experiments in which a few particles (approximately 100)  are introduced in the plasma and these are seen to be levitated at the plasma sheath boundary due to a balance between the gravitational and the electrostatic forces acting on them.  Typically, depending upon the discharge parameters, the dust particles float about a cm above the grounded {tray} and form a single layer.  In order to characterise this dusty plasma, we adjust the pumping speed and the gas flow rate in such a way that we can achieve an equilibrium steady state dust cloud.  To study the dynamics of the particles, a series of images of dimension 1000 pixels $\times$ 1000 pixels are captured into a computer from the CCD cameras. A sample image of such a collection of dust particles is shown in Fig.~\ref{fig:fig5}(a).  
%%%%%%%%%%%%%%%%%%%%%%%%%%%%%%%
%%%%%%%%%%%%%%%%%%%%%%%%%%%%%%%
An idl based sPIT, (super) Particle Identification and Tracking, code \cite{uwe,feng} is used to track individual particles over a large number of frames.  By using this code we can extract the coordinates of all the particles very precisely for a given span of time. By knowing the coordinates  of each particle for a single frame (e.g. for Fig.~\ref{fig:fig5}(a)), it is very straightforward to evaluate the pair correlation function which is shown in Fig.~\ref{fig:fig5}(b). The pair correlation function represents the probability of finding two particles separated by a distance r. It is generated by measuring the distance from each particle to every other particle and then counting the number of particles in the region between  r and $r+\delta r$ from the particle. This is repeated for every particle until an average value is determined which is then normalised by dividing by the annular area between r and $r+\delta$r. As we can see in Fig.~\ref{fig:fig5}(b) there is only one primary peak or a primary peak followed by a small second or third peak in the plot of the correlation function. In the present case the nature of the correlation function shows the existence of short range ordering between the particles which is indicative of the system being in a liquid state \cite{smith}. The first peak of the Fig.~\ref{fig:fig5}(b) shows the first nearest neighbour distance, from which we can estimate the  inter-particle distance. We can further estimate the dust density by considering a sphere of radius $r$ and counting the number of particles in volume $4\pi r^3/3$. From the above considerations, the inter-particle distance and density comes out to be $\sim 500 \mu$m and $\sim 2\times10^9 /m^3$ respectively.\par
The thermal velocity of  the particles can also be estimated by dividing the displacement by time of the same particles in consecutive frames (the time interval of two consecutive frames is known). In our case, for better statistics, we have chosen approximately a hundred particles for a few hundred frames to calculate the velocity distribution function.  Fig.~\ref{fig:fig5}(c) and Fig.~\ref{fig:fig5}(d) show a velocity distribution of dust particles along the x and y directions for discharge parameters of $V_d=290$ volt and $P=0.14$ mbar. A Maxwellian function is then fitted (solid line) on the experimental data points to estimate the temperature (from the full width at half maximum of the distribution function) of the dust particles. The dust temperature varies from  0.6-1.5eV along the x direction and 0.4-1.4eV along the y direction over a wide range of discharge parameters.
{It may be mentioned here that for most of our experimental conditions the inter-particle distance of $\sim 500 \mu$m is large compared to the Debye screening length of $\sim 30-50 \mu$m. Hence their motion is not likely to impact the motion of nearby particles. However when experiments are performed at lower pressures (and/or lower voltage) the radial confinement becomes stronger and as a result the dust particles can come closer to each other. In such a situation the particle motion can be affected by the other particles and this would need to be accounted for.}
%%%%%%%%%%%%%%%%%%%%%%%%%%%
\begin{figure}[hb]
\includegraphics[width=0.48\textwidth]{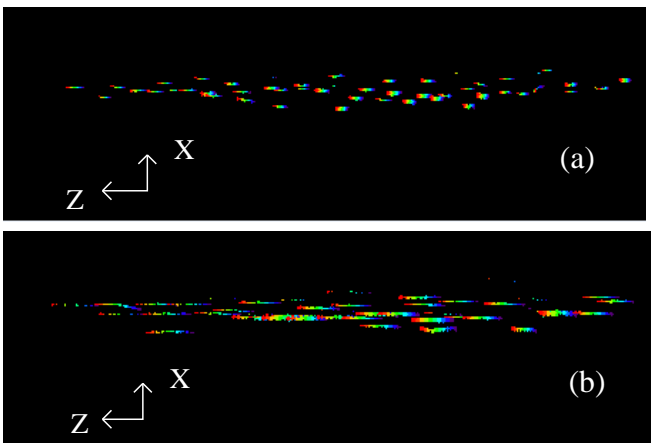}
\caption{Trajectory of particles for difference of gas flow rate (a) 2.75 m$l_s$/min and (b) 5.5 m$l_s$/min. The violet points represent the initial position whereas the red point represent the final position.} 
\label{fig:fig6} 
\end{figure}
%%%%%%%%%%%%%%%%%%%%%%%%%%%
\section{Experimental control of particle flow}
We now discuss the method of introducing equilibrium flows in the dust particles and the means of measuring and controlling such flows. As mentioned above, a steady state equilibrium dust cloud can be formed when the pumping rate and the gas flow rate are suitably balanced in a precise manner. {For a very short duration (less than a second)}, if the pumping rate exceeds the gas flow rate, the particles are seen to flow from right to left and in the reverse direction if the pumping rate is reduced below the gas flow rate. {By doing so we disturb the equilibrium condition momentarily. However, this gradient of pressure is created for less than one second and the maximum instantaneous change of pressure (as measured in the pressure meter) is less that $\pm 5 \%$. In fact sometimes this change cannot even be detected by the pressure meter if the flow rate is decreased by a small amount. Soon after the gas feed flow rate is set to its original value the equilibrium condition is restored.  To cross-check we connect the pressure gauge near the gas feeding port which also shows the same results. The basic mechanism responsible for the flow of the dust particle is the neutral flow acting on them from the flowing gas particle. With this change of sudden gas flow, the particles are initially accelerated towards the pump from the equilibrium condition. After travelling a distance of less than 1 cms,  almost all the particles are found to achieve terminal velocities, which is approximately equal to the background neutral velocity \cite{markus}. The state of terminal velocity is achieved in about 200 msec due to the combined action of the neutral drag force and the ion drag force (which acts in the opposite direction {of flow}). However, in most discharge conditions the neutral gas density is $10^5$ to $10^6$ times more than that of ions and hence the neutral drag is the dominating force. To ascertain this we have estimated the ratio of the ion drag force to the neutral drag force\cite{nakamura2} and found the neutral drag force to be always $\sim 10^3-10^4$ times higher than the ion drag force. A similar estimation has also been provided by Nakamura \textit{et al.}\cite{nakamura2}.} {In our experiments the neutral gas} flow is directly governed by the rate of pumping and gas injection rate. To characterize this flow we have made a series of measurements by systematically reducing the gas flow from the equilibrium conditions while keeping the pumping rate constant. {It is worth mentioning that we did not observe any flow vortices or other complex gas flow structures in the video images of the dust particles indicating that the flow is of a laminar nature.} \par
Fig.~\ref{fig:fig6} shows the trajectories of particles for discharge parameters, $V_d=300$ Volt and $P=0.130$ mbar {after attaining the terminal velocity}. The figures are drawn by overlapping 10 consecutive frames by different colours in the colour sequence of rainbow. The violet colour corresponds to the initial position of the particles and red colour corresponds to the final position of the particles.  From such a sequence it is clear that the particles are moving from right to left. Fig.~\ref{fig:fig6}(a) shows results obtained when the gas flow rate is changed from $27.5$ m$l_s$/min to $24.75$ m$l_s$/min, while 
~\ref{fig:fig6}(b) displays data for the case when the flow rate is changed from 27.5 m$l_s$/min to 22 m$l_s$/min.  It is clearly seen that the trajectories are shorter in Fig.~\ref{fig:fig6}(a), in comparison to the trajectories of particles shown in  Fig.~\ref{fig:fig6}(b), which implies that the velocity of the particles increases with the increase of flow rate. \par
%################## ################################################# 
%%%%%%%%%%%%%%%%%%%%%%%%%%%
\begin{figure}[ht]
\includegraphics[width=0.45\textwidth]{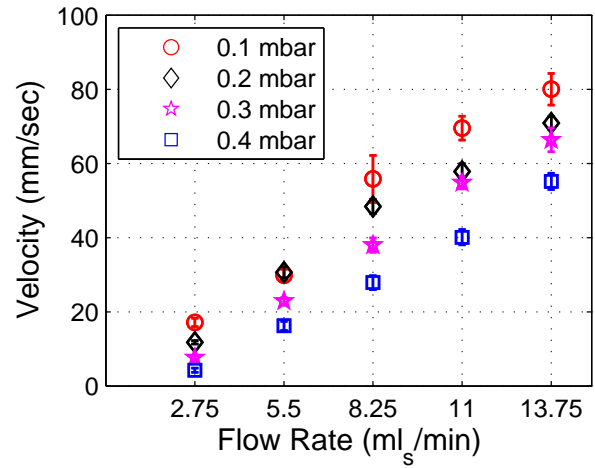}
\caption{Variation of particle velocity with the difference of gas flow rate.} 
\label{fig:fig7} 
\end{figure}
%%%%%%%%%%%%%%%%%%%%%%%%%%%
Fig.~\ref{fig:fig7} shows the variation of the average particle velocity with the change of gas flow rate from the equilibrium conditions for four different background pressures. {The  average velocity is plotted in this figure when the particles achieve their terminal velocities. It is to be noted that the pressure mentioned here is the pressure just before disturbing the equilibrium condition}. For a given value of neutral gas pressure ($P=0.110$ mbar), the velocity increases almost linearly with the change of gas flow rate from the equilibrium condition and establishes a reliable method for the control of dust flow in the device.  It is also worth mentioning that for a particular change of gas flow rate, the average particle velocity decreases with the increase of background gas pressure. It suggests that the particles get slowed down after making more collisions with the neutral gas molecules at higher pressure.\par
{As discussed earlier, the neutral drag force always dominates over the ion drag force for our discharge conditions and is responsible for slowing down the dust particle acceleration and bringing them to the terminal velocity. We now provide an estimate of this force using the measured flow velocities of the particles as plotted in Fig.~\ref{fig:fig7}. Assuming the particles to have a uniform acceleration ($\frac{dv}{dt}$) before attaining terminal velocity, the accelerating force acting on the particles can be expressed as, $F_n=m_d\frac{dv}{dt}$ as has been discussed in similar experiments performed by Thomas \textit{et al.} \cite{ed} (here, $m_d$ represents the mass of the particles). When the particles acquire terminal velocity this force can be equated with the neutral drag force given by $-\frac{4}{3}\gamma_{Eps}{\pi}r_d^2m_nN_nv_{thn}v_r$ \cite{nakamura2,ed}, where, $m_n$, $N_n$, $v_{thn}$ and $v_n$ are the mass, background density, thermal and drift velocities of the neutrals respectively. $\gamma_{Eps}$ represents the Epstein drag coefficient.  For a given value of discharge voltage $V_d=310$ volt, the neutral drag force is calculated over a wide range of background pressures used in Fig.~\ref{fig:fig7}. In that range of pressures,  the neutral drag force varies from $1.0\times10^{-13}$N to $4.0\times10^{-14}$N  when the gas flow rate is changed to 2.75 ml$_s$/min from equilibrium. From the measured value of $v_r$ we can calculate the Epstein drag coefficient which comes out to be approximately  $1.2-1.0$ for the above mentioned pressure range.} \par
These measurements and the primary technique can prove useful for conducting basic experiments on flow induced instabilities in dusty plasmas on DPEx.\par  
%##############################################################################3  CONCLUSION
\section{Conclusion}
\label{sec:conclusion}
A newly built table top experimental device, named DPEx, for the study of dusty plasma physics is presented. One of the unique features of the device is its ability to induce a flow in the dust component in a controlled fashion by adjusting the vacuum pumping speed and the gas flow rate of the device. This will facilitate an experimental study of flow induced instabilities and the associated formation of nonlinear structures in a dusty plasma - an area of research that is relatively unexplored till date. The geometry and construction of the device also facilitates viewing and making optical measurements of the dynamics of the dusty plasma. The device has been commissioned and is operational at the Institute for Plasma Research. Initial experiments to characterise the plasma and the dusty plasma have been performed and the results are reported here. Experimental results on the observation of flow of the dust component have also been obtained and it is shown that the particle velocity increases linearly with the increase of the gas flow rate but decreases with the increase of the neutral gas pressure. {Estimates of the neutral drag force and the Epstein drag coefficient are provided for a specific dusty plasma condition}. We intend to carry out a series of experiments on this device related to the excitation of wake fields, solitons and other nonlinear structures in the presence of various amounts of flow and these results will be presented in future publications. \\
\section*{Acknowledgements}
{The authors thank  Minsha Shah for her technical help in developing the Langmuir and emmisive probe diagnostics systems.}
%+++++++++++++++++++++++++++++++++++++++++
%\section*{References}
%###################################################################################

\end{document}